\documentclass[12pt,preprint]{aastex}
\usepackage{epsf}

\tighten

\newcommand{\gtsim}{\ {\raise-0.5ex\hbox{$\buildrel>\over\sim$}}\
}
\newcommand{\ltsim}{\ {\raise-0.5ex\hbox{$\buildrel<\over\sim$}}\
}
\def \etal{{\it et~al.~}}

\def\simlt{\lower.5ex\hbox{$\; \buildrel < \over \sim \;$}}
\def\simgt{\lower.5ex\hbox{$\; \buildrel > \over \sim \;$}}

\shortauthors{NEWMAN ET AL.}
\shorttitle{Measuring the Cosmic Equation of State}

\begin{document}

\title{{Measuring the Cosmic Equation of State with Galaxy Clusters in the DEEP2 Redshift Survey}}

\author{Jeffrey A. Newman, Christian Marinoni, Alison L. Coil, and Marc Davis\altaffilmark{1}} 

\affil{Department of Astronomy, University of California, Berkeley, CA 94720-3411}
\email{jnewman@astro.berkeley.edu, marinoni@astro.berkeley.edu, acoil@astro.berkeley.edu, marc@astro.berkeley.edu}

\altaffiltext{1}{Also Department of Physics, U.C. Berkeley}
\vskip -12pt
\begin{abstract}

The abundances of groups and clusters of galaxies are expected to have
changed substantially since high redshift, with the strength of this
evolution dependent upon fundamental cosmological parameters.
Upcoming large redshift surveys of distant galaxies will make it
possible to measure these quantities at $z\sim 1$; when combined with
the results of local redshift surveys currently underway, the
evolution of cluster abundances may be determined.  The DEEP2 Redshift
Survey, planned to begin in Spring 2002, is particularly well-suited
for this work because of the high spectroscopic resolution to be used;
velocity dispersions of groups will be
readily measurable.  In this paper, we determine the constraints on
dark energy models that counts of clusters within the DEEP2 survey
should provide.  The velocity function of clusters may be predicted
directly in the extended Press-Schechter framework.  We find that
comparing cosmological models using the simultaneous distribution of
clusters in both velocity dispersion and redshift yields significantly
stronger constraints than the redshift distribution alone.  The method
can be made more powerful by employing a value of the fluctuation
amplitude $\sigma_8$ determined with upcoming techniques (external to
DEEP2) that have no $\Omega_m$ degeneracy.  The equation-of-state
parameter for dark energy models, $w=P/\rho$, can then be measured to
$\pm 0.1$ from observations of clusters alone.

\end{abstract}

\keywords{cosmological parameters, cosmology: observations, galaxies: high-redshift, galaxies: clusters: general}

\section{Introduction}

In two previous papers (Newman \& Davis 2000, hereafter ND00, and
Newman \& Davis 2001, hereafter ND01), we described a new variant of
the classical ``d$N$/d$z$'' test that could measure fundamental
cosmological parameters using data from the next generation of
redshift surveys.  By measuring the apparent abundance of galaxies
as a function of their linewidth or velocity dispersion rather than
luminosity or other properties, we may exploit the simplicity of the
velocity function of dark matter halos and avoid many of the
uncertainties that result from the physics of galaxy formation.
Combining measurements of the velocity function at low and high
redshift yields the evolution of the cosmic volume element, which
depends upon fundamental cosmological parameters in a simple fashion.
This is not the only form of the d$N$/d$z$ test to be considered in
recent years, however.

In particular, because clusters of galaxies are rare, particularly at
high redshift, their number density is exponentially sensitive to the
rate of growth of large-scale structure.  Their observed abundance
thus can place limits on fundamental cosmological parameters (Lilje
1992).  For instance, Haiman, Mohr, \& Holder (2001) found that the
observed numbers and redshift distributions of galaxy clusters
discovered in future X-ray or Sunyaev-Zel'dovich (S-Z) surveys could
impose strong constraints on the cosmic equation-of-state parameter of
quintessence-like dark energy models, $w= P /\rho$.  However, because
it relies on counting the total number of clusters above some minimum
mass, a rapidly decreasing function, their method requires the mass
limit for finding clusters in such surveys to be very well-understood.
They find that the mapping between the matter power spectrum and the
masses of clusters leads to a further dependence on the Hubble
parameter, the value of which is still only known to $\sim 10\%$
(Freedman \etal 2001).

In this paper, we propose another form of this test which will be
possible using data from the same galaxy redshift surveys as the
method of ND00.  Just as it is possible to count galaxies as a function
of their circular velocity rather than their optical luminosity, we
can count galaxy clusters as a function of their velocity dispersion
rather than their X-ray luminosity or S-Z decrement.  By performing the
test on the differential velocity function rather than the integrated
count above some mass, we do not require perfect knowledge of the
survey characteristics to produce results.  Furthermore, by
studying clusters as a function of their velocity dispersions rather
than their masses, we can avoid the sensitivity to the Hubble parameter
present in other methods.

Specifically, we present here the constraints upon cosmological
parameters which the DEIMOS/DEEP (hereafter, DEEP2) Redshift Survey
will provide (Davis \etal 2000).  This project is intended to obtain
data on large samples of distant galaxies using the new DEIMOS
spectrograph, which is scheduled to be installed in early 2002 at the
Keck Observatory.  DEEP2 will obtain spectra of $\sim 60,000$ galaxies
preselected from $BRI$ photometry to have minimum redshift $z>0.7$
(the ``1HS'', or 1-Hour Survey, so named because of the expected
exposure time per slitmask).  Four 2 deg $\times$ 1/2 deg fields have
been selected for the 1HS, yielding a total volume (for the optimal
redshift range of the survey, $0.7 < z < 1.5$) approaching $10^7
h^{-3}$ Mpc$^3$ in LCDM cosmologies.  The 1HS will have a magnitude
limit of $I_{AB}=23.5$, roughly $L_{*}$ at $z=1$.  Roughly 70\% of
the galaxies meeting the survey criteria will be targetable for
observations (due to the technical constraints of slitmask
spectroscopy), and secure redshifts are expected for $\sim 85\%$ of
the observed galaxies.  In addition, longer-exposure spectra of $\sim
5,000$ galaxies to $I_{AB}=24.5$ will be obtained in selected regions
(roughly 10\% of the total survey area), constituting the ``3HS'', or
3-Hour Survey.  DEEP2 will obtain data characterizing galaxies and
large-scale structure that are comparable to those provided by the
best completed surveys of the local universe, but for objects at high
redshift, $z\sim1$, instead.  Because of the high spectroscopic
resolution (FWHM $\sim 65$ km s$^{-1}$ at $z=1$) and relatively dense
sampling to be used, DEEP2 will be uniquely suitable for providing
measurements of the velocity dispersions of galaxy clusters in the
distant universe with no preselection.  Other past or planned projects
such as the VLT/VIRMOS survey and its subsamples (Lefevre 2000) only
have sufficient redshift resolution to determine the velocity
dispersions of only the most massive clusters, only cover small areas
of the sky, and/or are likely to be less densely sampled than DEEP2 at
$z\sim 1$ because of their shallower magnitude limits and lack of
selection against low-redshift objects.  As an additional advantage,
sensitive S-Z observations are planned for all DEEP2 fields, which
will allow the virialization state of clusters found to be assessed.
In $\S$ 2 of this paper, we describe our
calculations of cluster abundances, and in $\S$ 4, the resulting
constraints upon fundamental cosmological parameters.

\section{Calculations of cluster abundances}

Narayan \& White (1988) showed, under the assumption that structures
observed are well-described by isothermal spheres and are just
virializing, that the velocity dispersion distribution of dark matter
halos may be calculated within the Press-Schechter (1974) framework as
simply as the mass distribution.  For this work, we apply their
technique to the improved semianalytic mass function of Sheth \&
Tormen (1999), using the approximate relations of Bryan \& Norman
(1996; for those models with $w=-1$) or Wang \& Steinhardt (1998) to determine
the overdensity of collapsed structures compared to the background
density, $\Delta_{vir}$.  We have fixed the power spectrum shape
parameter $\Gamma=0.25$ in our calculations.  Given a value of the
fluctuation normalization $\sigma_8$, the velocity dispersion
distribution of dark matter halos in a given cosmology follows
immediately; we assume here that the measured velocity dispersions of
the galaxies within clusters will follow the same distribution.  Since
clusters even today are dynamically very young, these assumptions are
expected to work very well, and indeed are borne out in comparisons to
N-body models (Springel \etal 2000).

In the following analysis, we consider two scenarios for the
determination of the mass power spectrum normalization $\sigma_8$.  In
one, which we will label as ``conservative'', we assume that studies
of galaxies and clusters in upcoming local surveys (such as 2dF and
SDSS, Colless 1998, Loveday \etal 1998) will fix the power spectrum
sufficiently that errors in cosmological parameters will be dominated
by cosmic variance and Poisson statistics in the DEEP2 sample, but
with the same parameter degeneracies that have affected past
measurements.  Since these surveys are much larger in volume and have
higher sampling density than DEEP2, this is likely to be the case.
Thus, in this scenario, for each cosmological model considered we use
the results of Borgani et al. 1999 (in cases where we have fixed
$w=-1$) or Wang \& Steinhardt (1998) to assign values for $\sigma_8$
as $\Omega_m, \Omega_{\Lambda}$, and $w$ vary, with zero error
assumed.

In the other scenario, which we will term ``optimistic,'' we presume
that emerging techniques which fix $\sigma_8$ for the mass with no
dependence on other cosmological parameters will be successful.  For
example, the 2dF and SDSS surveys will provide extremely accurate
measurements of the correlation properties of nearby galaxies.  Weak
lensing analyses, by measuring either the mass in individual galaxy
halos (e.g. McKay \etal 2001) or of the aggregate large-scale
structure (Kaiser 1998) can then determine the bias between the
correlation statistics of galaxies and of the underlying dark matter,
allowing transformations from one to the other.  From preliminary SDSS
data, for instance, McKay \etal found that in the red optical bands
($r$, $i$, and $z$), the light of nearby galaxies traces the mass on
scales up to 1 Mpc, and that the influence of groups is clear.  If
weighted by luminosity in these bands, galaxy correlation measurements
should then provide an accurate estimator of the mass correlation
function, and thus of $\sigma_8$.  Measurements on non-linear scales
may be reliably connected to the equivalent linear amplitude using the
methods of Hamilton \etal (1991).  If $\sigma_8$ has been determined
from such external data, we can use the abundances and velocity functions
of local clusters in conjunction with those at high redshift to set
better constraints on cosmological parameters.

It is necessary to note that the Press-Schechter framework upon which
these calculations are built relies upon the Gaussianity of
fluctuations in the matter density.  If that fundamental assumption
fails, the abundances of clusters at high redshift, which lie on the
extreme tail of the probability distribution for density, may differ
radically from Gaussian predictions.  In that case, the observed
abundances of clusters in DEEP2 will place few constraints on
cosmological parameters, if any, but could provide strong information
on cosmological non-Gaussianity (Robinson \& Baker 2000).

\section{Cosmological constraints}

Given the methods for calculating the abundance of clusters described
in $\S 2$, we may compare the predictions of various models for the
observed number of clusters per unit redshift and solid angle to
determine what constraints on cosmological parameters will be possible
from DEEP2.  This may either be done integrally (comparing the total
number of clusters above some velocity dispersion observed in
different redshift intervals, d$N(>\sigma)/$d$z$, to the predictions
for a model) or differentially (using the distribution of clusters in
velocity dispersion as well as redshift, d$N/$d$\sigma$d$z$, to set
constraints).  We have therefore calculated the comoving abundance of
clusters over dense grids in velocity dispersion, redshift, and
cosmological parameters ($\Omega_m$ and $\Omega_{\Lambda}$ for models
with $w=-1$, or $\Omega_m$ and $w$ for models assumed to be flat).
The grid spacings used are sufficient to allow determination of the
integrated abundance of clusters in ten 50 km s$^{-1}$ velocity
dispersion bins from 300 to 800 km s$^{-1}$ (along with an eleventh
bin for clusters with velocity dispersions from 800 to 1000 km
s$^{-1}$, beyond which very few objects are predicted to exist) and in
8 bins spanning $z=$0.7 to 1.5, each covering 0.1 in redshift.  The
results may then be multiplied by the amount of volume in each
redshift bin for the DEEP2 survey in the given cosmology to yield a
prediction for the observed number of clusters in each
bin.  For the optimistic scenario, we have also calculated the
expected observed abundance of clusters in each model for a survey
spanning $0<z<0.1$ covering one-fourth of the sky (using the velocity
function at $z=0.05$), similar to what one might expect for the
densely sampled portion of the SDSS Redshift Survey (Loveday \etal
1998).

Poisson variance should be the dominant source of uncertainty in a
measurement of the abundance of clusters with DEEP2.  In an LCDM
model, less than a thousand clusters with velocity dispersions above
$400$ km s$^{-1}$ are expected to exist in the survey volume;
considerably fewer should be found in any of the redshift/velocity
bins.  For the lowest-velocity, most abundant clusters, a comparable
error may arise from cosmic variance, the excess fluctuations in
counts of cosmological objects that occur because of large-scale
correlations.  Unlike Poisson variance, this uncertainty will be
correlated in every velocity bin within the same redshift interval.
We have used here the DEEP2 cosmic variance calculations of ND01
rescaled to a value of $\sigma_8=1.8$, which matches the amplitude of
fluctuations of 400 km s$^{-1}$ clusters at $z\sim 1$ expected from
their predicted correlation length in an LCDM model (Colberg \etal
2000).

Any application of the cluster d$N$/d$z$ test may be subject to a
variety of systematic effects: the identification of clusters and the
measurement of their velocity dispersions in an unbiased way are
inherently difficult, even at low redshift (see, for instance,
Giuricin \etal 2000).  Clusters are not actually the isothermal
spheres we have assumed in the velocity function predictions, nor do
galaxies precisely trace the dark matter potential of clusters.
However, in actually performing a d$N$/d$z$ measurement, one can be
guided by comparisons to the results of N-body simulations, in which
clusters may be found and counted with the same systematics that
affect DEEP2 observations, instead of using simple semi-analytic
predictions.  We thus believe this is not likely to be a crippling
problem.  So long as the measurement errors (or the errors due to any
systematics) in the velocity dispersions of the clusters are known
from theory or tests with simulations, those errors may be applied to
the predictions of each cosmological model before comparison to
observations.  With sufficient theoretical effort towards determining
the relationship between the observed properties of clusters and their
intrinsic characteristics, the constraints presented here could be
achieved; we focus on the limits of what will be possible with the
data.  We have reason to be optimistic; Marinoni \etal (2001) find
that new cluster identification and membership determination
algorithms, when applied to mock DEEP2 catalogs drawn from the
VIRGO/GIF simulations enhanced with semianalytic techniques, can
reconstruct the actual cluster velocity function from observations down to a
velocity dispersion of 300 km s$^{-1}$.

For clusters found in a large local survey, cosmic variance is
negligible; the volume considered is orders of magnitude higher than
that in any DEEP2 redshift bin.  The number of clusters in all but the
most extreme velocity bins will accordingly be large as well.  We thus
may expect that systematic errors are more likely to dominate over
Poisson errors than they are at high redshift.  For constraints at low
$z$, we therefore have conservatively required the uncertainty
assigned to the abundance in each redshift/velocity bin to be at least
5\%, with the Poisson value used if it is larger than that.

Given the above definitions, the covariance matrix amongst the
redshift and velocity bins is fully determined (as Poisson variance is
uncorrelated in both redshift and velocity, while the cosmic variance
yields a completely correlated fractional error amongst velocity bins
at the same redshift, but is nearly uncorrelated between different
bins of 0.1 in $z$).  We may then calculate $\chi^2$ between any model
and some nominal, ``true'' model (e.g. LCDM: $\Omega_m=0.3$,
$\Omega_Q=0.7$, $w=-1$) in either the conservative or the optimistic
scenario.\footnote{We use the extension of $\chi^2$ to a multivariate
distribution with covariance: $\chi^2=(\bf{n-n_0})^T V^{-1}
(\bf{n-n_0})$, where $\bf{n}$ is the vector of observations,
$\bf{n_0}$ is the vector of true values, and $\bf{V}$ is the
covariance matrix for $\bf{n_0}$.}  Observed results should be
distributed as $\chi^2$ with two degrees of freedom, so contours of $\chi^2$ may be immediately transformed into statistical confidence constraints.

\begin{figure}
\begin{footnotesize}
\renewcommand{\baselinestretch}{1.0}
\epsscale{.9}
\plotone{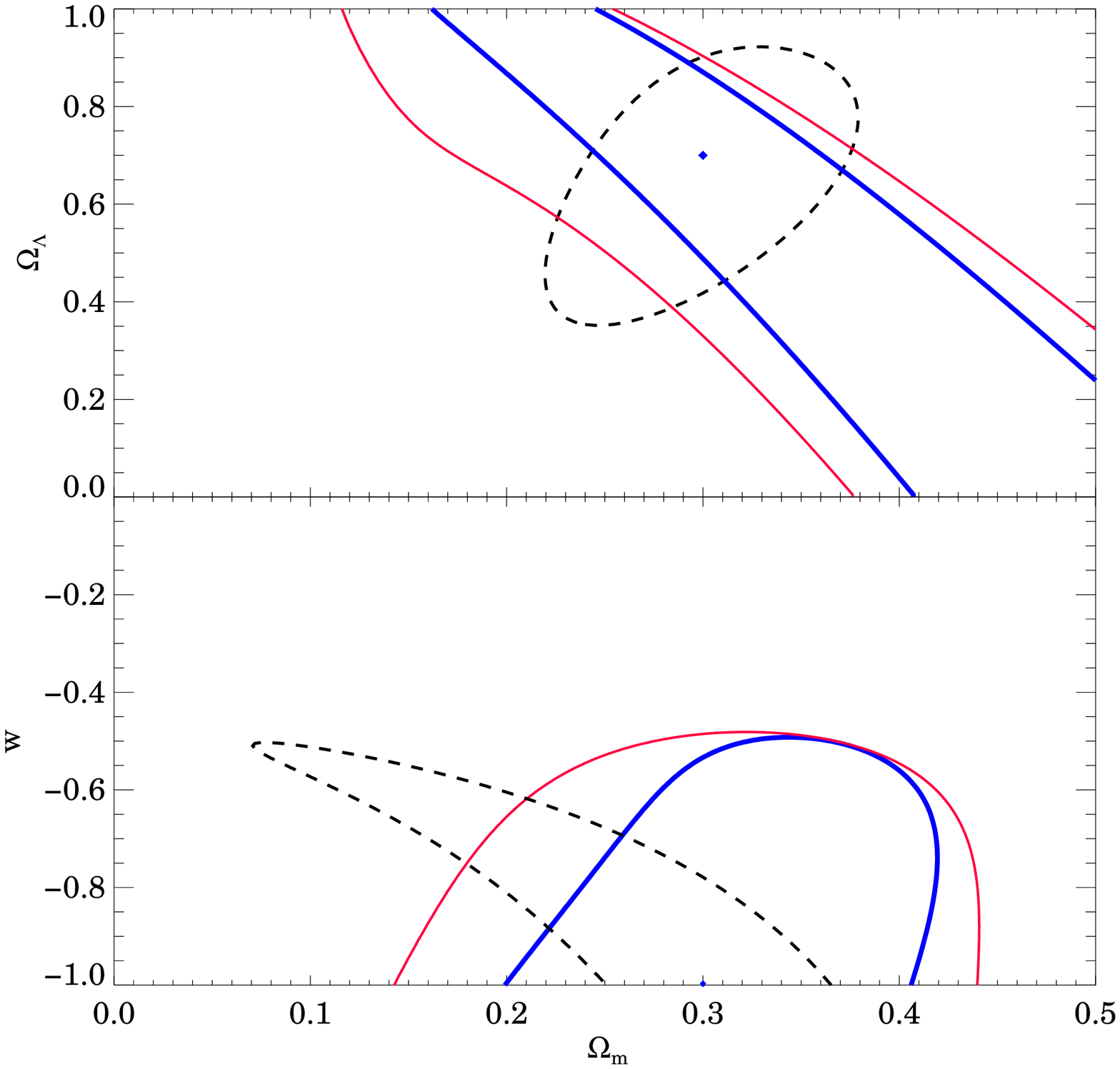}
\caption{Expected constraints from a variety of cosmological tests
made possible by the DEEP2 Redshift Survey, plotted for an LCDM model
with $\Omega_m=0.3$, $\Omega_{\Lambda}=0.7$, and $w=-1$ (indicated on
all plots by a small blue diamond).  All contours are at the 95\%
confidence level.  (Top panel) Constraints in the
$\Omega_m$--$\Omega_{\Lambda}$ plane.  The black, dashed contour is
the ``best bet'' constraint from galaxy d$N$/d$z$ measurements (see
ND01 for details).  We also plot two sorts of constraints from
DEEP2 clusters above 400 km s$^{-1}$ in the conservative scenario: the
thick, blue contours show the results of utilizing the distribution of
those clusters in both velocity dispersion and redshift,
d$N/$d$\sigma$d$z$, while the thin, red solid contours use only the
integrated number of clusters above 400 km s$^{-1}$ in each redshift
bin, ignoring all differential information on the velocity function.
(Bottom panel) As above, but for the $\Omega_m$--$w$ plane. (NOTE:
When printed out on a black and white printer, ``red'' contours in
these figures will appear grey while the ``blue'' contours will be
nearly black.  In that case, the line style may still be used to
distinguish the curves).
\label{comboc1} }
\end{footnotesize}
\end{figure}

\begin{figure}
\begin{footnotesize}
\renewcommand{\baselinestretch}{1.0}
\epsscale{.9}
\plotone{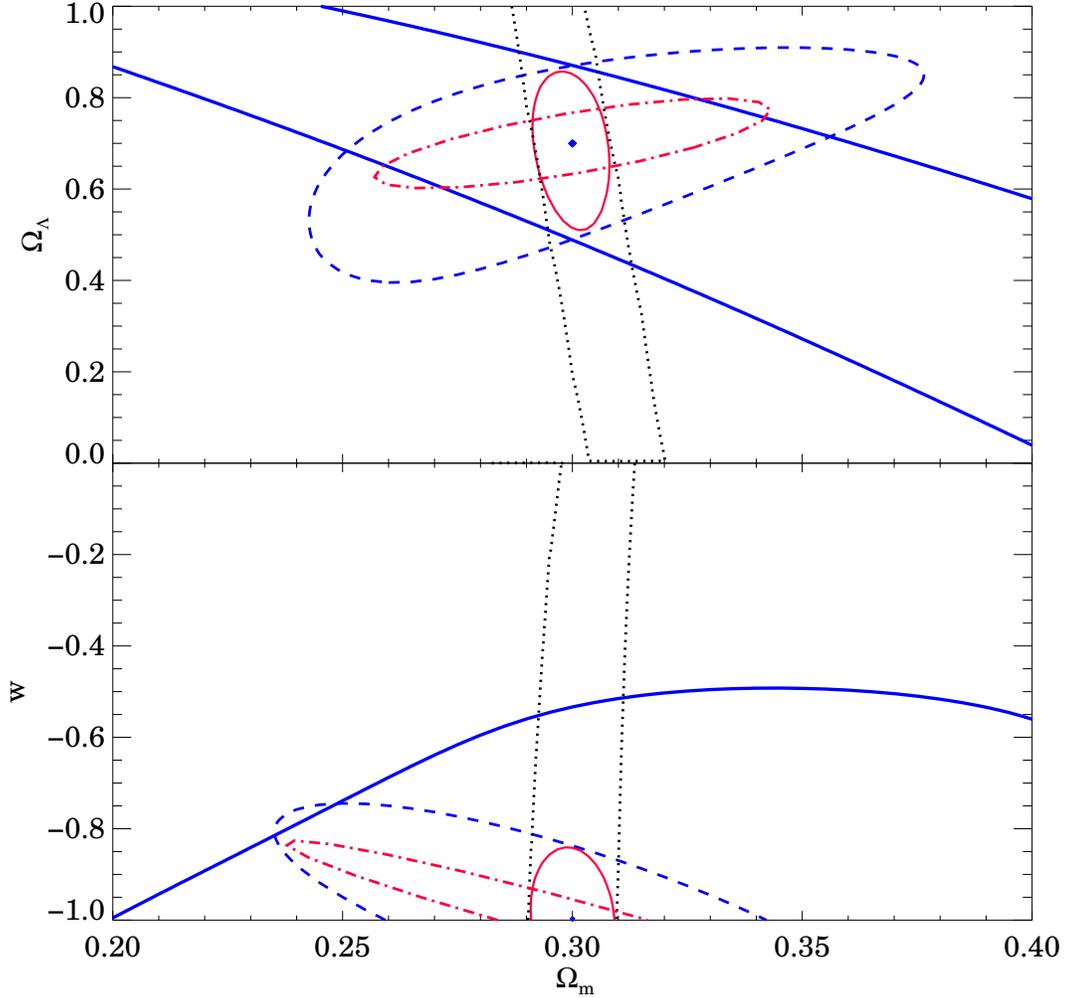}
\caption{Cluster constraints on cosmological parameters obtainable
using a variety of methods in an LCDM model.  All contours are at the
95\% confidence level.  (Top panel) Constraints in the
$\Omega_m$--$\Omega_{\Lambda}$ plane (note that a more restricted
range in $\Omega_m$ is plotted than in Fig. \ref{comboc1}).  As in the
previous figure, the thick, blue contours show the constraints
obtainable from d$N/$d$\sigma$d$z$ in the conservative scenario.  The
dashed blue curves show the constraints from DEEP2 if the value of
$\sigma_8$ for the mass is known with no other parameter degeneracies,
while the dotted black contours show the constraints an SDSS-like survey
could then provide from the velocity function of low-redshift
clusters.  The thin, solid red contours represent the optimistic
scenario in which data from low redshift and high redshift may be used
simultaneously to strengthen the constraints.  The red, dot-dashed
curves show for comparison the target 95\% confidence intervals
(statistical errors only) for the proposed SNAP satellite, a dedicated
orbiting telescope to find SNe Ia at high redshift, taken from figures
on the project's website (http://snap.lbl.gov).
(Bottom panel) As above, but for the $\Omega_m$--$w$
plane. \label{comboc2} }
\end{footnotesize}
\end{figure}

\begin{figure}
\begin{footnotesize}
\renewcommand{\baselinestretch}{1.0}
\epsscale{1}
\plotone{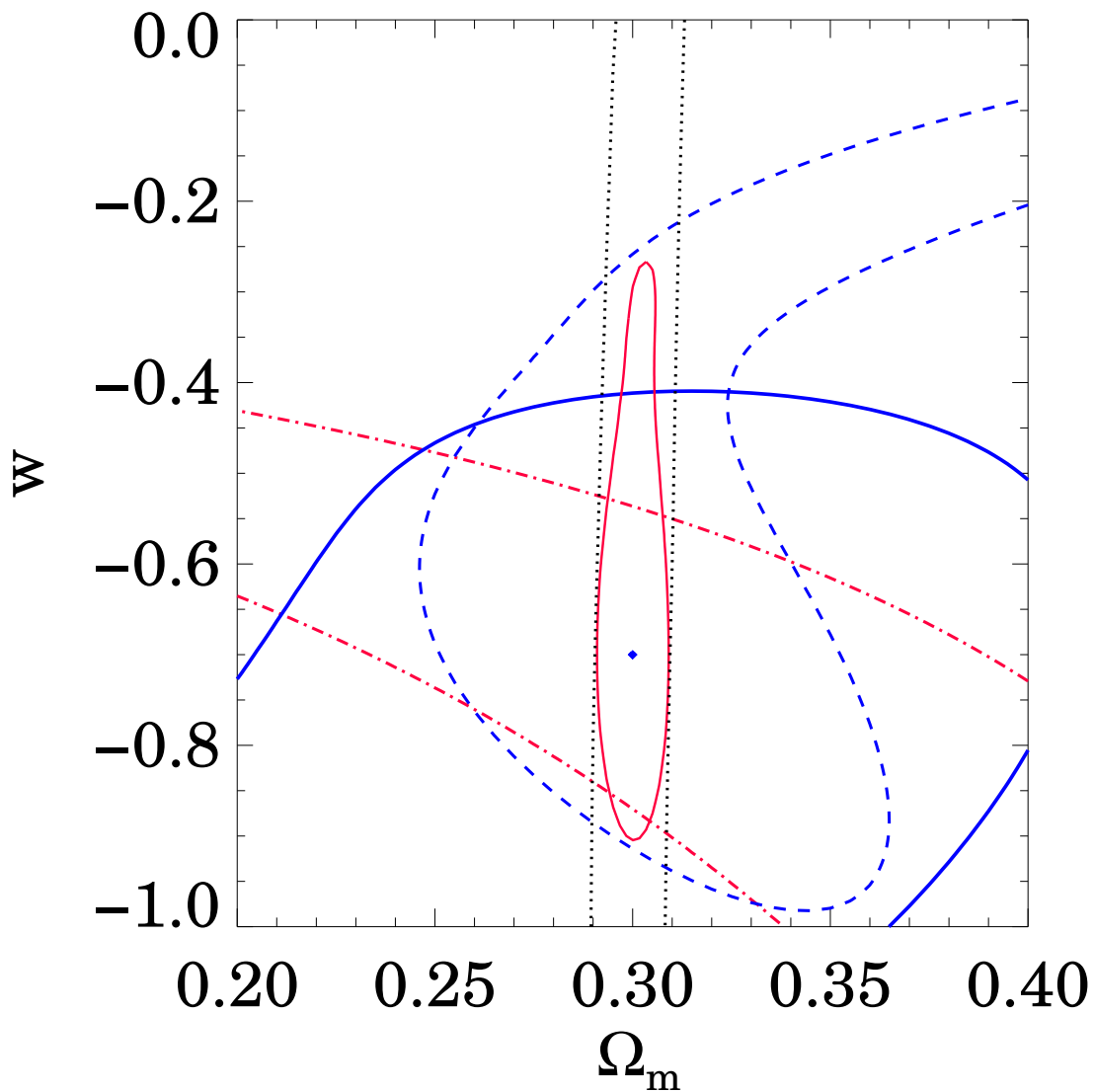}
\caption{Cluster constraints on cosmological parameters obtainable
using a variety of methods in a quintessence model with $\Omega_m=0.3,
\Omega_Q=0.7, w=-0.7$.  The solid, dashed, and dotted contours are
all defined as in the preceding figure.  The red, dot-dashed contour
indicates the ``best bet'' constraint from DEEP2 galaxy d$N$/d$z$
measurements for this model, taken from ND01.  \label{comboc72} }
\end{footnotesize}
\end{figure}

In Fig. \ref{comboc1} we show the results of these calculations for
the conservative scenario, assuming that clusters may be reliably
found down to a velocity dispersion of 400 km s$^{-1}$ (the actual
limits will depend upon our ability to identify and measure the
characteristics of small clusters; see Marinoni \etal 2001).
Measuring the distribution of clusters in both velocity dispersion and
redshift rather than using only d$N(>\sigma)/$d$z$ yields
substantially stronger parameter measurements.  We have also plotted
in this figure the ``best bet'' contours from ND01.  This method is
subject to completely different systematic effects, providing an
excellent consistency check.  As shown in Fig. \ref{comboc2}, using
the optimistic $\sigma_8$ normalization yields much stronger
constraints than any of those presented in Fig. \ref{comboc1},
especially if $z\sim 0$ information is used.  In that case, the value
of $w$ may be determined to better than 10\% from cluster observations
alone.  We have also plotted for comparison the target 95\% contours
(statistical errors only) for observations of 2000 distant SNe Ia by
the SNAP satellite (Perlmutter \etal 2000).  A precision determination
of $\Omega_m$ such as that obtained in the optimistic scenario would
be highly complementary to the SNAP observations, yielding much
stronger constraints on cosmological parameters; in the absence of a
precision measurement of $\Omega_m$, SNAP and DEEP2 cluster
constraints on $w$ would be quite comparable, but with very different
systematics. Fig. \ref{comboc72} depicts the constraints set by DEEP2
for a model with $w=-0.7$.  As is true for many methods
(e.g. observations of SNe Ia at high redshift; see Huterer \& Turner
2001), cluster d$N/$d$\sigma$d$z$ observations yield much weaker
constraints on $w$ if its value is not -1; however, DEEP2 galaxy
d$N$/d$z$ observations provide a very useful complementary constraint,
yielding in combination a measurement of $w$ to $\sim10\%$.

\begin{figure}
\plotone{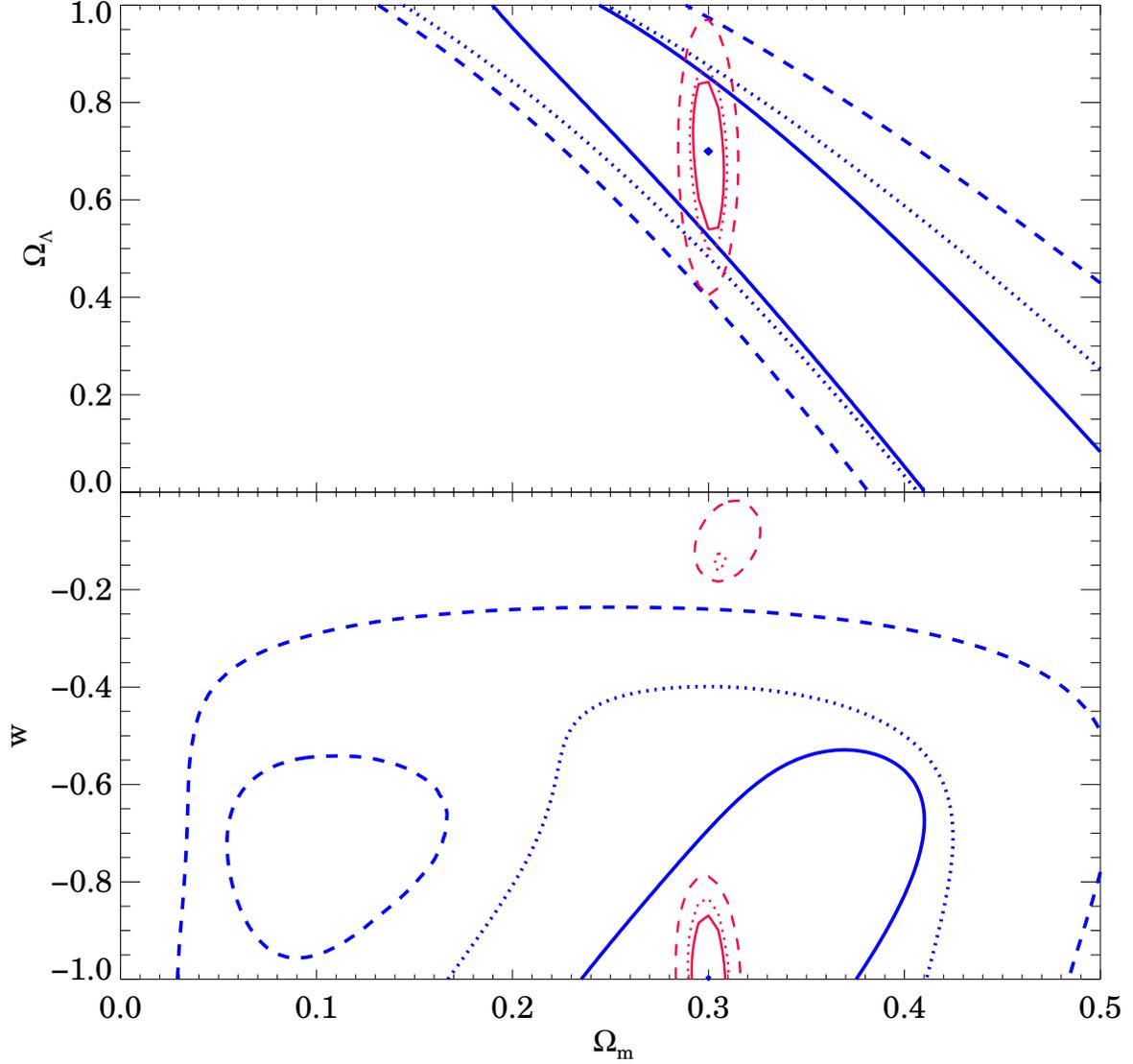}
\epsscale{.9}
\renewcommand{\baselinestretch}{1.0}
\caption{Variation of constraints plotted in Fig. \ref{comboc2} as the
minimum velocity dispersion is changed.  The blue contours show the
DEEP2 constraints for an LCDM model in the conservative scenario,
while the red contours indicate the combined low- and high-redshift
constraints for the optimistic scenario.  The solid, most optimistic
contours are for the case where all velocity bins of dispersion $>300$
km s$^{-1}$ and above are used; the dashed, intermediate contours
$>500$ km s$^{-1}$; and the dot-dashed, weakest contours $>700$ km
s$^{-1}$.  The two panels are defined as in Fig. \ref{comboc1}.  The
degenerate solutions in the bottom panel with $w>-0.2$ are ruled out
by other cosmological tests (Huterer \& Turner 2001).  Note that there
is an excluded region within the conservative, $\sigma > 700$ km
s$^{-1}$ contour.  Paradigms which simplify parameter constraints
(such as Fisher matrix methods) completely fail to describe the
contours for large minimum velocities.
\label{vminc1} }
\end{figure}

Fig. \ref{vminc1} shows the dependence of the constraints upon the
minimum velocity dispersion measured.  In the optimistic scenario
where low and high redshift clusters are studied, the constraints are
nearly identical if only clusters above 500 km s$^{-1}$ are considered
as if clusters are observed down to a dispersion of 300 km s$^{-1}$.
Although at $z \sim 1$ there are only $\sim 10\%$ as many clusters
above 500 km s$^{-1}$ as above 300 km s$^{-1}$ ($\sim 300$ versus
$\sim 3000$ in an LCDM model), even in the conservative scenario the
constraints are only modestly weaker.  Because of the strong
dependence of their abundance upon the rate of growth of structure,
the largest, rarest clusters have a weight in determining cosmological
parameters that is disproportionate to their abundance.
Large, local surveys should be very effective for finding these extreme
clusters.
On the other hand, the volume surveyed by DEEP2 is sufficiently small
that only $\sim 10$ clusters above 800 km s$^{-1}$ velocity dispersion
will be observed, so it is less possible to exploit their exponential
sensitivity to the growth of structure from DEEP2 alone.  Although
they will be unable to detect the smaller DEEP2 clusters and groups,
upcoming S-Z experiments will be capable of finding massive objects
over much larger areas, $\sim 1000$ deg$^2$ (Holzapfel 2001).  With
suitable follow-up observations, they S-Z results could be used to
tighten further the potential constraints obtained from the velocity
dispersion and redshift distributions of DEEP2 clusters presented
here.

\begin{figure}
\plotone{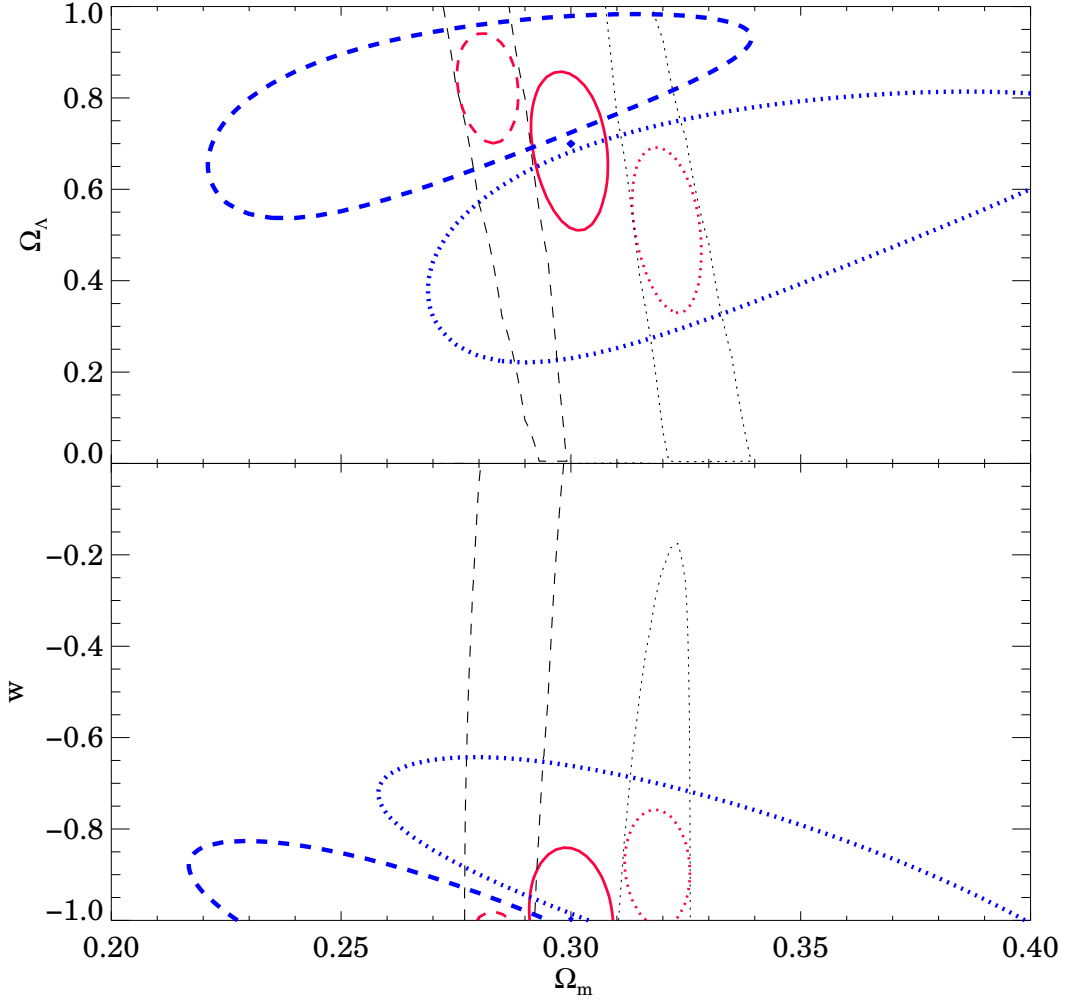}
\epsscale{.9}
\renewcommand{\baselinestretch}{1.0}
\caption{The variation of the DEEP2 cluster d$N$/d$z$ constraints for an
LCDM scenario if incorrect, fixed values of $\sigma_8$ are used (e.g.,
there is some systematic error in determining the bias, and thus also
in $\sigma_8$ for the mass).  The thin, black contours show
constraints from $z \sim 0$ clusters alone; the thick, blue contours
constraints from $z \sim 1$ clusters; and the red contours the
combined constraints.  For the dotted curves, a value of $\sigma_8$
that is too low by 5\% has been used to determine constraints; for the
dashed curvess, a value that is too high by 5\%; and for the solid
curves (for clarity, only plotted in the case of the combined
constraint), the true value has been used.  This figure covers the
same restricted range of $\Omega_m$ depicted in Fig. \ref{comboc2}.
\label{sigmac1} }
\end{figure}

Even if the value of $\sigma_8$ used in the optimistic scenario is
uncertain, useful constraints on cosmological parameters may be
obtained.  In Fig. \ref{sigmac1}, we show the results of an error in
$\sigma_8$ of $\pm 5\%$.  As would be expected from previous parameter
measurements based upon local clusters (e.g Borgani et al. 1999), if
an erroneous value of $\sigma_8$ is used, a statistically equivalent
distribution of local clusters may still be obtained for some (also
erroneous) value of $\Omega_m$.  However, the high-redshift contours
respond very differently; the primary change to the combined
constraint is an offset of $\sim 10\%$ in the best-fit values of
$\Omega_m$ and $\Omega_{\Lambda}$ or $w$.  The increased precision
afforded by the optimistic normalization makes the sensitivity of the
measurement to the determination of $\sigma_8$ of equal or even
greater importance than statistical errors.

In conclusion, we find that counts of clusters observed in the DEEP2
Redshift Survey have the potential to provide significant constraints
on cosmological parameters, particularly when combined with both a
non-cluster constraint on $\sigma_8$ and measurements of the local
cluster velocity function.  The data have sufficient power that the
utility of this test is likely to be limited by our theoretical
understanding and simulation capabilities rather than the
observations.  DEEP2 cluster constraints can be complementary to a
variety of other tests that have been proposed, including not only
studies of SNe Ia at high redshift or counts of clusters found by
their S-Z decrement, but also galaxy d$N$/d$z$ measurements that the DEEP2
survey will make possible.  By comparing and combining the results of
very different methods of determining cosmological parameters, we may
both obtain stronger constraints than any method alone would provide
and test techniques against each other to identify signatures of
systematic effects.  In a field so afflicted by systematic errors as
cosmology, having many complementary techniques is the best way to
ensure that our framework of measurement holds together.

\acknowledgements

We would like to acknowledge useful conversations with Andrew Jaffe,
Proty Wu, and especially Martin White.  This material is based upon
work supported by the National Science Foundation under Grant
No. AST-0071048.  This work was also made possible by equipment
donated by Sun Microsystems.


\end{document}